\documentclass[journal]{IEEEtran}
\usepackage{amsmath,amsfonts}
\usepackage{algorithmic}
\usepackage{algorithm}
\usepackage{array}
\usepackage{textcomp}
\usepackage{subfigure}
\usepackage{stfloats}
\usepackage{url}
\usepackage{verbatim}
\usepackage{graphicx}
\usepackage{cite}
\usepackage{xcolor}
\newcommand{\ml}[1]{{\color{black} #1}}

\begin{document}

\title{Safety Analysis for Distributed Coupled-Cavity Laser based Wireless Power Transfer}

\long\def\symbolfootnote[#1]#2{\begingroup%
\def\thefootnote{\fnsymbol{footnote}}\footnote[#1]{#2}\endgroup}
\renewcommand{\thefootnote}{\fnsymbol{footnote}}
\author{Mingqing~Liu,~\emph{Member, IEEE,} Hao~Deng,~\emph{Member, IEEE,} Iman Tavakkolnia, \emph{Member, IEEE,}\\ Qingwen~Liu,~\emph{Senior Member, IEEE,}~Bin He,~\emph{Senior Member, IEEE,}~and~Harald Haas, \emph{Fellow, IEEE}

%
\thanks{
\emph{This work is supported in part by the Engineering and Physical Sciences Research Council (EPSRC) under grant EP/X040569/1 ‘Future Communications Hub in All-Spectrum Connectivity’.}

\emph{Mingqing Liu and Bin He are with the College of Electronics and Information Engineering, Tongji University, Shanghai, China (email: \{clare, hebin\}@tongji.edu.cn). Mingqing Liu was working with LiFi Research and Development Centre, Electrical Engineering Division, University of Cambridge, Cambridge, UK. Hao Deng and Qingwen Liu are with the School of Computer Science and Technology, Tongji University, Shanghai, People's Republic of China (e-mail: \{denghao1984,qliu\}@tongji.edu.cn). Iman Tavakkolnia~and~Harald Haas are working with LiFi Research and Development Centre, Electrical Engineering Division, University of Cambridge, Cambridge, UK (email:\{it360, huh21\}@cam.ac.uk). }
}}


\maketitle

\begin{abstract}
Intracavity laser-based systems are emerging as key enablers for next-generation wireless communications, positioning, and wireless power transfer (WPT). \ml{Distributed coupled-cavity laser (DCCL) systems, as a representative configuration, have been proposed to expand the field of view (FoV) and enhance safety.} This paper investigates the safety assessment of DCCL-WPT systems through three case studies: skin safety, eye safety, and small-object intrusion sensitivity. First, we establish a safety analysis model \ml{to quantify irradiation levels on intruding objects in the beam path, which simulates intracavity beam propagation using diffraction modeling and gain-loss dynamics under case-specific boundary conditions.} 
Next, we formulate an eye safety evaluation tailored for DCCL-WPT systems using a human head model to identify potential exposure angles and distances. Ray tracing confirms that intracavity beams are not focused onto the retina, making cornea exposure the primary consideration (irradiance is below $0.1$ W/cm$^2$). Numerical results \ml{demonstrate that DCCL-WPT achieves: i) over $600$~mW charging power under skin-safe conditions at $5$ m distance ($100$ mW over $16^{\circ}$ FoV), and nearly $50$\% lower irradiance on intruding objects compared to single-cavity systems; ii) $150$~mW charging power under eye-safe conditions with $650$ mW $1064$ nm output beam power, far beyond the typical $\sim 10$ mW eye-safe threshold; iii) high sensitivity to small-object intrusion, enabling hazard mitigation.} These findings underscore the practicality of DCCL-WPT systems for mobile, long-distance, and safe energy transfer, and lay the groundwork for future safety-aware optimizations in real-world deployments.
\end{abstract}


\IEEEpeerreviewmaketitle

\section{Introduction}
\label{sec:Introduction}
Laser technologies offer compelling advantages for wireless power transfer (WPT), including high directivity, broad modulation bandwidth, and immunity to electromagnetic interference, making them especially suitable for long-distance and electromagnetically constrained environments~\cite{jin2018wireless,hosseinvcsel,wen2025reservoir}. Although laser-based communication, typically operating in the near-infrared band, has achieved impressive data rates and high directionality~\cite{LIFI2,laserpositioning}, extending its application to WPT remains constrained by stringent eye safety requirements and mobility limitations that restrict deliverable power~\cite{soltani2022safety}. To address this, intracavity laser architectures have emerged with self-alignment and inherent safety properties, and the recently proposed distributed coupled-cavity laser (DCCL) design further expands alignment tolerance and improves safety~\cite{basicMobility}. However, the eye safety of the DCCL has not yet been systematically investigated. This paper fills that gap by proposing a comprehensive safety analysis framework for DCCL-based WPT (DCCL-WPT) systems, covering skin exposure, eye safety, and sensitivity to small-object intrusions.

\ml{According to laser safety standards such as IEC 60825-1-2014, laser radiation hazards are primarily categorized into photochemical effects (dominant below $780$ nm) and thermal effects (dominant above $780$ nm)~\cite{schulmeister2013upcoming}. For infrared lasers, such as the $1064$ nm or $980$ nm wavelength range that have been experimentally demonstrated in intracavity systems due to their high electro-optical conversion efficiency~\cite{tianjin1,vecselExp,wwExpe}, the main safety concern is thermal damage due to high power density. In this context, the maximum permissible exposure (MPE) serves as a key metric in evaluating human safety, especially for skin and eyes.} Skin exposure to laser radiation is typically considered safe up to $1$ W/cm$^2$~\cite{jean2019comparison}. However, \ml{the eye is far more vulnerable due to its lens focusing effect, which can concentrate laser energy onto the retina~\cite{eyeMPE}.} As a result, the MPE for eye exposure is set much lower, typically at $5$ mW/cm$^2$. Although wavelengths beyond $1400$ nm are not focused onto the retina and are considered relatively safer, near-infrared lasers at $1064$ nm, still present significant eye safety risks~\cite{soltani2022safety}, \ml{and thus must be carefully assessed when designing user-centric intracavity systems}.

\ml{Prior research on intracavity laser systems has focused on distributed single-cavity lasers (DSCL), in which the transmitter comprises a gain medium and a retroreflector (RR), while the receiver contains a second RR to complete one resonant cavity~\cite{YUN1064nm,ShengSingle}. Owing to their dual-RR cavity design and the dynamic interplay between gain and loss, DSCL can automatically establish resonant links even as there is an angle between transceivers~\cite{mobilityEnhanced,mobilitySMIPT}. When an object intrudes into the cavity (i.e., free-space transmission path), the increased loss immediately disrupts the transmission, ensuring an inherent safety mechanism. Still, if intruding objects such as human body parts, inadvertently enter the intracavity path, the irradiance they receive during the brief transient before beam termination remains a safety concern. Following this idea, a theoretical model has demonstrated that the irradiance on the surface of an intruding object in DSCL, throughout the entire process until the beam is cut off, is only $1/10$ of that in traditional laser systems. The results indicate that DSCL can meet skin-safety requirements with $2.5$ W output optical power under static transceiver conditions~\cite{basicsafety}. However, the tight coupling between gain medium and free-space path in DSCL leads to high intracavity power densities, posing challenges for compliance with stricter safety standards such as eye safety. Moreover, DSCL architectures face intrinsic trade-offs between system gain and FOV. These limitations motivate growing interest in DCCL designs~\cite{tianjin1,basicMobility,SHENGDCCL}. Integrating photon-to-electricity converters such as photodetectors (PDs) or photovoltaic (PV) cells can extend the functionality of DCCL toward WPT or communication applications.}

\ml{In DCCL-WPT, the gain medium and two internal RRs are fully integrated at the transmitter, forming a main cavity. Another RR at the transmitter and one at the receiver jointly form a separate free-space resonant cavity, which is optically coupled to the main cavity. Meanwhile, beam power is output from the receiver side and converted by PV. DCCL separates the gain mechanism from both the FoV constraints and the need for high free-space power density, aiming to maintain real-time disruption upon object intrusion while simultaneously reducing the maximum surface irradiance and expanding the FoV. A mobility analysis model has demonstrated that DCCL can expand the FoV from $\pm 10^{\circ}$ in DSCL to $\pm 40^{\circ}$\cite{basicMobility}. As for safety, an optical semiconductor amplifier-based prototype has shown that the resonating power in free space reaches $17.2$ mW under alignment and drops by a factor of four when obstructed, suggesting the potential for eye-safe operation\cite{lim58wireless}. Inspired by DCCL, protective beam designs have also been proposed, surrounding the high-power core with low-power shielding beams, though their effectiveness in ensuring eye safety remains to be validated~\cite{safetyprotect}.
Quantitative assessment of DCCL safety improvements and their compliance with eye safety standards is a key issue to be addressed.}

\ml{Safety analysis of DCCL-WPT can draw on existing DSCL skin-safety models, where intrusion is simulated by modeling the large-scale object as a knife-edge gradually approaching the optical axis~\cite{basicsafety}.} As the intrusion deepens, the irradiance on the object's surface initially increases until the intracavity laser ceases oscillation due to excessive cavity loss. \ml{Beyond skin safety, eye safety evaluation must account for the fact that the human eye is anatomically embedded within the head, rather than being an isolated organ.} This raises a unique safety mechanism: as the head approaches the intracavity beam, cranial features such as the forehead or brow ridge are likely to intersect the beam path before the eye itself is exposed. To investigate this hypothesis, we construct a refined safety model that incorporates the human head morphology and examines whether the intracavity beam is physically blocked or defocused before reaching the retina. This assessment further evaluates the feasibility of relaxing the conventional retinal MPE limit ($5$ mW/cm$^2$) to the corneal exposure threshold ($100$ mW/cm$^2$)~\cite{corneaMPE}. On these bases, we extend the safety analysis to DCCL-WPT, and conduct following quantitative analysis: i) the improvement in skin safety relative to DSCL configurations, ii) the validation of human eye safety, and iii) the sensitivity of DCCL to small-object intrusion.

Our contributions in this manuscript are summarized as 
\begin{itemize}
\item[i)]  We present a comprehensive safety evaluation framework for DCCL-WPT systems, enabling quantitative analysis under various mobility and intrusion scenarios. Numerical results show that DCCL-WPT achieves over $600$ mW of charging power at $5$ m under skin-safe conditions (maintaining over $100$ mW within a $16^{\circ}$ self-alignment FoV), while reducing irradiance on large-scale intrusions by nearly $50$\% compared to single-cavity systems.

\item[ii)]
We propose an eye safety analysis methodology tailored for DCCL-WPT, combining anatomical modeling and ray tracing. By showing that facial structures often block the beam path and that the intracavity laser is not focused onto the retina, we relax the evaluation from retinal MPE to corneal-level exposure. Results confirm that with $650$ mW output at $1064$ nm (typical threshold is $\sim 10$ mW), DCCL-WPT remains below the $0.1$ W/cm$^2$ corneal limit, achieving $150$ mW eye-safe charging power.

\end{itemize}

The remainder of this paper is organized as follows. Section II introduces the safety analysis framework for DCCL-WPT systems, incorporating modeling of intruding objects and charging power conversion. Section III presents the eye safety assessment approach based on human head modeling, beam focusing simulation, and irradiance analysis. Section IV provides comprehensive numerical evaluations demonstrating the enhanced skin safety, eye safety, and small-object intrusion sensitivity of DCCL-WPT systems. Finally, Section V concludes the paper and discusses directions for future work.

\begin{figure*}
    \centering
    \includegraphics[width=5.5in]{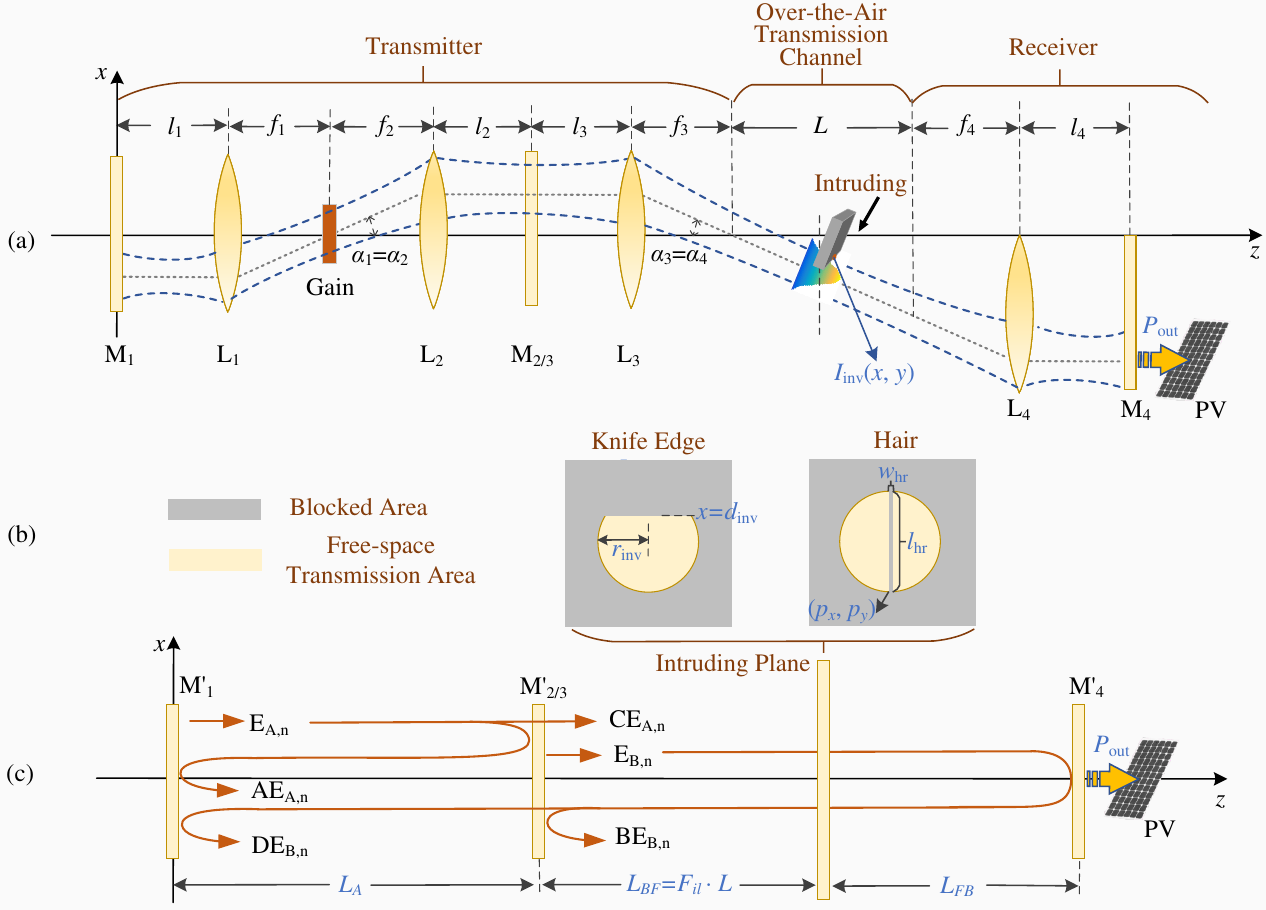}
    \caption{Safety analysis framework for DCCL-WPT systems: (a) Coordinate system used for DCCL-WPT safety evaluation; (b) Boundary condition definitions for different types of intruding objects; (c) Beam propagation schematic in the presence of an intruding object.}
    \label{fig:overview}
\end{figure*}

\section{Safety Assessment for Foreign Intrusions}
\ml{In DCCL systems, laser beams circulate continuously within a free-space resonant cavity through round-trip propagation. Unlike conventional laser systems that face the risk of direct beam exposure to skin or eyes, the primary safety concern in DCCL systems arises when an object physically intrudes into the intracavity beam path before beam termination. In such cases, evaluating the irradiance on the surface of the intruding object is essential for safety assessment. This section begins by establishing a safety assessment framework with intruding objects, where the key task is to model the beam field under boundary-defined diffraction conditions and simulate the gain-loss dynamics to determine the resulting surface irradiance. The system is considered safe only if this irradiance, denoted as $I_{\rm inv}$, remains below the MPE threshold $I_{\rm MPE}$. Finally, the wireless charging power converted by PV is derived.

\subsection{System Overview}
This work aims to evaluate safety across various intrusion scenarios, including both large-scale objects (e.g., human body parts) and small-scale elements (e.g., hair). An exemplary DCCL system is adopted as in Fig.~\ref{fig:overview} (a), consisting of four cat's eye retroreflectors (CRRs). Each CRR$_i (i={1,2,3,4})$ compromises a lens L$_i$ with focal length $f_i$ and a mirror M$_i$ positioned at a distance $l_i$ from the lens. The first two CRRs form the main cavity, housing the gain medium, while the other two create the free-space cavity, i.e., over-the-air transmission channel, where objects may intrude. The $z$-axis passes through the centers of all components on the transmitter side and is normal to their surfaces. The receiver, located at a distance $L$ from the transmitter, may deviate from the $z$-axis, causing the optical axis between the transceivers to tilt. As a result, the incident angles of the beam on each CRR are denoted as $\alpha_i$, respectively. The proposed DCCL-WPT system operates based on intracavity laser generation within a coupled resonator structure. When the input power excites the gain medium, laser oscillation can be initiated as long as the coupled cavity formed by CRRs supports round-trip beam propagation without excessive optical leakage, so that laser beam undergoes repeated amplification while passing the gain medium multiple times. Energy coupling is achieved through M$_{2/3}$. At the receiver, mirror M$_4$ possesses a defined transmittance, allowing part of the stabilized intracavity laser to be extracted and directly projected onto a PV panel. This emitted beam is then converted into electrical power through PV conversion, enabling safe and mobile WPT.

The intruding object is assumed to enter the cavity along a direction perpendicular to the optical axis. In beam propagation simulations, interaction surfaces are typically represented using indicator functions: regions where the beam can propagate are assigned a value of $1$, surrounded by $0$ values to prevent edge bias in numerical calculations~\cite{FFT}. The intruding plane is modeled using such an indicator function, where the intruding object is further represented as an opaque region (assigned value $0$) that blocks the intracavity laser beam. Examples of both large-scale and small-scale intruding objects are illustrated in Fig.~\ref{fig:overview}(b). Let $r_{\rm inv}$ denote the radius of the free-space transmission area on the intruding plane. This region is typically much larger than the beam spot or any optical aperture in the system, allowing us to neglect aperture-induced diffraction and treat it as ideal free space.
For large-scale objects, a knife-edge model is employed~\cite{hodgson2005laser}, i.e., as the object intrudes further, it progressively occludes the free-space transmission area. The invasion depth is denoted as $x = d_{\rm inv}$, representing the position of the object along the $x$-axis. For small-scale objects, such as hair, we model the obstruction as a rectangular opaque strip with length $l_{\rm hr}$ and width $w_{\rm hr}$. 

To simulate beam propagation within a DCCL-WPT system under mobility conditions, an equivalent geometric model is adopted~\cite{basicMobility}. In this model, the original cavity with CRRs and varying transceivers angles is approximated by a cavity formed by flat mirrors aligned in parallel, as shown in Fig.~\ref{fig:overview}(c). Each CRR$_i$ is equivalently replaced by a flat mirror $M_i'$, with the effective reflection area adjusted based on the beam incident angle. To describe the horizontal position of the intrusion, we define the fill factor $F_{il}$, such that the distance between the transmitter-side mirror ($M_{2/3}'$) and the intruding plane is given by $L_{BF} = F_{il} L$. Since the intracavity laser beam undergoes continuous bidirectional propagation within the free-space resonant cavity, we denote by $I^+$ the irradiance on the surface of the intruding object due to forward (transmitter-to-receiver) transmission, and by $I^-$ the irradiance due to backward (receiver-to-transmitter) transmission. The total  irradiance on the intruding object surface is given by}
\begin{equation}
    I_{\rm inv} = I^+ + I^-,
\end{equation}
where $I^{\pm}$ can be obtained by beam intensity distribution $I^{\pm}(x,y)$ on intruding object's surface. Therefore, the safety analysis for the intruding object relies on calculating the beam field intensity on the intruding object. 

\subsection{Key Planes for Beam Propagation Analysis}

To determine the beam field intensity in DCCL-WPT, a self-consistent equation describing the round-trip propagation of intracavity lasers must be established~\cite{FFT}.
\ml{A critical first step is to define the key planes where the beam passes or interacts within the system, primarily consisting of the intruding plane and the planes corresponding to the CRRs. These planes serve as the boundary conditions for the self-consistent equation. The mathematical formulation of these boundary conditions in the DCCL-WPT is presented below.}

\subsubsection{Intruding Plane with Knife-Edge} The knife edge area is set much larger than the CRRs of the resonant cavity to indicate a large intruding object like a human hand, head, etc. As shown in Fig.~\ref{fig:overview} (b), given the radius of the intruding plane $r_{\rm inv}$ and the intruding position $x = d_{\rm kni}$, the effective surface of the large-scale intruding object is described as~\cite{basicsafety}
\begin{equation}
    T_{\rm kni}(x, y)=\left\{\begin{aligned}
&1, x^2+y^2 \leq r_{\rm inv}^2 \& x \leq d_{\rm kni} \\
&0, x^2+y^2>r_{\rm inv}^2 \& x>d_{\rm kni}
\end{aligned}\right..
\end{equation}
The above equation assumes that the knife edge invades along $x$-axis, and along $y$-axis in the same manner.

\subsubsection{Intruding Plane with Hair} Hair can be simplified as a rectangle with length indicating hair length and width indicating hair diameter. Therefore, we denote $l_{\rm hr}$ and $w_{\rm hr}$ as length and width, respectively, and $(p_{x}, p_y)$ as the bottom left coordinate of the hair that intrudes into the resonant cavity. Then, the effective surface of the intruding hair is given by
\begin{equation}
    T_{\rm hr}(x, y)=\left\{\begin{aligned}
&1, x^2+y^2 \leq r_{\rm inv}^2  \\
&0, x^2+y^2>r_{\rm inv}^2 \\
&0, x \in[p_x, p_x+w_{\rm hr}]  \&  y\in [p_y, p_y+l_{\rm hr}] 
\end{aligned}\right..
\label{eq:hairbound}
\end{equation}
\subsubsection{Equivalent Plane for CRR} \ml{As illustrated above, a resonant cavity composed of CRRs can be equivalently modeled as a cavity with flat mirrors, where the size of each mirror corresponds to the effective reflective area of the associated CRR under mobility conditions}~\cite{mobilitySMIPT,mobilityEnhanced}. In the exemplary DCCL-WPT design as in Fig.~\ref{fig:overview} (a), the effective reflection surface of CRR$_i$ with $l_i=f_i$ is given by~\cite{basicMobility}
\begin{equation}
T_i(x, y)= \begin{cases}1,&x^2+y^2 \leq r_{\rm c}^2 ~\&~ x^2+(y-a_i)^2 \leq r_{\rm c}^2 \\ 0, & \text {else }\end{cases}.
\end{equation}
Here, $r_{\rm c}$ is the radius of both the lens and mirror in CRR$_i$, and $(0, a_i)$ is the center of the shifted lens aperture, which is determined by the incident beam angle $\alpha_i$ to the CRR surface \ml{caused by angle between transceivers}, described as~\cite{basicMobility}
\begin{equation}
a_i = 2f_i\tan{\alpha_i}.
\end{equation}

\subsection{Beam Intensity on Intruding Object}
\ml{The beam field distribution is analyzed using the equivalent cavity structure illustrated in Fig.~\ref{fig:overview} (c). M$'i$ has amplitude reflectivity and transmission of $r_i$ and $t_i$ respectively, where $|r_i|^2+|t_i|^2=1$, neglecting losses.}

\subsubsection{Self-Consistent Equation} The guess field on M$'_1$\ml{, i.e., the
estimation of the field at a given iteration,} is $E_{A,n}$, which propagates to M$'_{2}$ where it is reflected and transmitted. The beam field from the free-space cavity also injects into the main cavity. The guess field on M$'_3$ for the free-space cavity is $E_{B,n}$, which propagates to M$'_4$ and part of it forms the output beam. Two iterative relations can be derived to find steady-state solutions for the main cavity and free-space cavity respectively by computing one round trip of the field with self-consistent equations~\cite{coupledcavityBeam}:
\begin{equation}
\begin{aligned}
E_{A, n+1}^{\prime}(x,y)&=\mathcal{D}\left[E_{B, n}(x,y)\right]+\mathcal{A}\left[E_{A, n}(x,y)\right] \\
E_{B, n+1}^{\prime}(x,y)&=\mathcal{B}\left[E_{B, n}(x,y)\right]+\mathcal{C}\left[E_{A, n}(x,y)\right]
\end{aligned},
\label{e:self}
\end{equation}
where $\mathcal{A}=\mathcal{R}_1 \mathcal{P}_A \mathcal{R}_2 \mathcal{P}_A$ is the operator describing the round-trip
propagation inside the main cavity, $\mathcal{B}=\mathcal{R}_3 \mathcal{P}_{BF} \mathcal{T}_{\rm inv} \mathcal{P}_{FB}\mathcal{R}_4 \mathcal{P}_{FB} \mathcal{T}_{\rm inv} \mathcal{P}_{BF}$ is the operator describing the round-trip propagation inside the free-space cavity, $\mathcal{C}=\mathcal{P}_A\mathcal{T}_2$ is the operator describing the one-way propagation from M$'_1$ to M$'_2$ and passes through M$'_2$, $\mathcal{D}=\mathcal{R}_1\mathcal{P}_A\mathcal{T}_2\mathcal{P}_{BF} \mathcal{T}_{\rm inv} \mathcal{P}_{FB}\mathcal{R}_4\mathcal{P}_{FB} \mathcal{T}_{\rm inv} \mathcal{P}_{BF}$ is the operator describing beam propagating from M$'_3$ to M$'_1$ as a right-hand field. Specifically, $\mathcal{R}_i$ and $\mathcal{T}_i$ for $i\in\{1,2,3,4\}$ are the reflection and transmission operators, respectively, defined as
\begin{equation}
\begin{aligned}
\mathcal{R}_i [E(x,y)] &= jr_iT_i(x,y)E(x,y)\\
\mathcal{T}_i [E(x,y)] &= t_iT_i(x,y)E(x,y)
\end{aligned},
\end{equation}
where multiplication by the imaginary unit $j$ indicates the reflection process of the mirror. $\mathcal{T}_{\rm inv}$ expresses the beam field passing through intruding plane as
\begin{equation}
\begin{aligned}
\mathcal{T}_{\rm inv} [E(x,y)]  = T_{\rm inv}(x,y)E(x,y),~{\rm inv} \in\{\rm kni, hr\}
\end{aligned}.
\end{equation}
Then, $\mathcal{P}_A$,  $\mathcal{P}_{BF}$ and $\mathcal{P}_{FB}$ are propagation operators for the main cavity, free-space cavity from transmitter to intruding object,  and free-space cavity from intruding object to the receiver, respectively, which are the solutions of the paraxial diffraction. We use a unified operator $\mathcal{P}_m, m\in\{A, BF, FB\}$ to express the corresponding operations as~\cite{FFT}
\begin{equation}
\begin{aligned}
&\mathcal{P}_m[E(x, y)]=\mathcal{F}^{-1}\left[H_m\left(f_x, f_y\right) \cdot \mathcal{F}\left[E(x, y)\right]\right]\\
&H_m\left(f_x, f_y\right)=\exp \left[j \frac{2\pi}{\lambda} L_m \sqrt{1-\left(\lambda f_x\right)^2-\left(\lambda f_y\right)^2}\right]
\end{aligned},
\end{equation}
\ml{where $H_m(f_x,f_y)$ is the propagation kernel as a function of the spatial frequency coordinates $(f_x,f_y)$.} In the given context, the fast Fourier transform (FFT) and inverse FFT (IFFT) are denoted by $\mathcal{F}$ and $\mathcal{F}^{-1}$, respectively. $L_m$ denotes the length of the main cavity, the distances between M$_{2/3}'$ to the intruding object, and the object to M$_4'$. $\lambda$ is the wavelength of the intracavity laser beam.

\subsubsection{Gain-Loss Interaction Simulation}
We incorporate gain process simulation into the self-consistent equation for beam propagation, where the beam field will be updated whenever it passes through the gain medium during each iteration~\cite{basicMobility}. Using rate equations to model the gain process, we can generalize the refreshment of the beam field $E$ by the gain medium in the simulation as~\cite{asoubar2016simulation}
\begin{equation}\left\{\begin{aligned}
    &E' =\left[\frac{I_{\mathrm{S}}}{c_0\epsilon_0\left|E\right|^2} W\left(\frac{c_0\epsilon_0}{I_{\mathrm{s}}} \exp \left(g_0 l_{\rm g}+c_1\right)\right)\right]^{\frac{1}{2}}\cdot E\\
    &c_1=\ln(2\left|E\right|^2)+\frac{c_0\epsilon_0\left|E\right|^2}{I_{\rm S}}, \quad g_0=\frac{\eta_{\rm c}P_{\rm in}}{I_{\rm S}V}
    \end{aligned}\right.,
\end{equation}
where $c_0=3\times 10^8$m/s is the light speed in the free space, $\epsilon_0=8.85\times10^{12}$F/m is the vacuum permittivity, $l_{\rm g}$ is the length of the gain medium, and $g_0$ is the small-gain coefficient which can be obtained by pump efficiency $\eta_{\rm c}$, input pump power $P_{\rm in}$, the volume of gain medium $V$, and $I_{\rm S}$~\cite{solid}. Above all, the relationship between the pump power introduced into the system, the amplification of beam power, and the beam propagation loss inside the cavity has been established.

\subsubsection{Output Power and Beam Intensity} Using the Fox-Li algorithm to iterate Eq.~\eqref{e:self}, both $E_{A,n}(x,y)$ and $E_{B,n}(x,y)$ reach a steady state where the residual for the intracavity laser stops changing after enough iterations. In the simulation, residuals represent the normalized difference in the sum of a field value between each step and are mathematically defined for a general intracavity beam field parameter $E$ as~\cite{convergence}
\begin{equation}
\Delta E^n=\frac{\sum_{p, q}|| E_{p, q}^n|-| E_{p,q}^{n-1}||}{\sum_{p,q}\left|E_{p,q}^{n-1}\right|},
\label{e:conve}
\end{equation}
where $p,q$ are the indices corresponding to the spatial grid, and $n$ is the simulation step, each corresponding to a simulated pass in the cavity. Finally, the beam field distribution converges according to Eq.~\eqref{e:conve}. Let the steady-state beam field on M$'_4$ be $E_4(x,y)$, we can then calculate the output laser power from M$'_4$ as follows~\cite{Pout}:
\begin{equation}\left\{\begin{aligned}
&P_{\rm out} = \iint_{x,y}\Gamma_4I_{4}(x,y){\rm d}x{\rm d}y,\\
&I_{4}(x,y) = \frac{1}{2}c_0\epsilon_0 \left|E_4(x,y)\right|^2
\end{aligned}\right.,
\label{eq:pout}
\end{equation}
where $\Gamma_4 = 1-r_4^2$ is the transmittance of M$'_4$.
\subsubsection{Irradiance on Intruding Plane} After obtaining the steady beam field $E_B(x,y)$ on M$'_3$ and $E_4(x,y)$ on M$'_4$, we compute the beam field on both sides of the intruding plane using the previously defined propagation operators as
\begin{equation}
\begin{aligned}
    E^+(x,y) &= \mathcal{T}_{\rm inv} \mathcal{P}_{BF} 
 [E_B(x,y)]\\
    E^-(x,y) &= \mathcal{T}_{\rm inv} \mathcal{P}_{FB} \mathcal{R}_{4} [E_4(x,y)]
\end{aligned}.
\end{equation}
The corresponding beam intensities, $I^+(x,y)$ and $I^-(x,y)$, are then derived from  Eq.~\eqref{eq:pout}. Finally, we determine the maximum irradiance on the intruding plane as $I^{\pm}$ using the $\max(\cdot)$ function, yielding $I_{\rm inv}$.

\subsection{Wireless Charging Power Conversion}
The laser beam output from M$_4'$ with optical power \( P_{\text{out}} \) is received by the PV cell and converted to electrical energy for battery charging. The wireless charging current \( I_{\text{chg}} \) is 

\begin{equation}
\begin{aligned}
    I_{\text{chg}} &= \rho P_{\text{out}} - I_0 \left[ \exp\left( \dfrac{qI_{\text{chg}} (R_{\text{L}} + R_{\rm s})}{n_s n kT} \right) - 1 \right]\\& - \dfrac{I_{\text{chg}} (R_{\text{L}} + R_{\rm s})}{R_{\text{sh}}},
\end{aligned}
\end{equation}
where \( \rho \) is the responsivity of the PV, \( I_0 \) is the reverse saturation current,  \( q \) is the elementary charge, \( n_s \) is the number of cells in the PV module, \( n \) is the ideality factor of the internal equivalent diode, \( k \) is the Boltzmann constant, and \( T \) is the absolute temperature in Kelivin. \( R_{\text{PL}} \), \( R_{\rm s} \), and \( R_{\text{sh}} \) represents the equivalent load resistance, internal equivalent series resistance, and internal equivalent shunt resistance, respectively. The charging power \( P_{\text{chg}} \), which is the output from the PV cell, also depends on the charging voltage \( V_{\text{chg}} \). In practice, a maximum power point tracking (MPPT) circuit is typically placed between the PV cell and the battery to optimize the power conversion efficiency. MPPT dynamically adjusts the effective load impedance \( R_L \) as seen by the PV source in order to operate at the maximum power point (MPP) on the current-voltage (I–V) characteristic curve. Finally, the output power from PV can be obtained by $P_{\rm chg} = I_{\rm chg}^2 R_{\rm L}$.

\section{Eye Safety Assessment Method}
When a laser beam directly enters the eye, it is focused by the eye's lens, significantly increasing its intensity and potentially damaging the retina. This is the primary reason why eye safety regulations are the most stringent at infrared wavelengths, with a typical MPE limit of $I_{\rm MPE}<5$mW/cm$^2$ to ensure retinal safety~\cite{eyeMPE}. However, intracavity lasers are highly sensitive to object intrusion, and in practical scenarios, the eyes are part of the human head. This suggests a potential mitigation mechanism: before the laser reaches the eyes, other facial structures may obstruct the beam, causing termination before it can be focused by the lens. In this case, the most hazardous scenario shifts to laser exposure on the cornea, where the MPE limit is significantly higher, at $I_{\rm MPE}<100$mW/cm$^2$~\cite{corneaMPE}.

A three-step validation process is employed to assess the proposed mechanism: i) determine the distance and angle between the eye and the first facial structure that interacts with the intracavity laser beam, which corresponds to the invasion depth of the head and the incident angle of the intracavity laser beam before reaching the eye; ii) establish an eye model and verify whether a laser beam entering at the distance and angle identified in Step i) would be focused onto the retina; iii) compute the irradiance on the obstructing structure as the intruding depth increases using models in Sec. II, ensuring that it remains below safety thresholds throughout the process until the beam is fully blocked and cutoff. 
\begin{figure}  
\centering
\subfigure[Human head model]{
\begin{minipage}[b]{0.22\textwidth}
\includegraphics[width=1\textwidth]{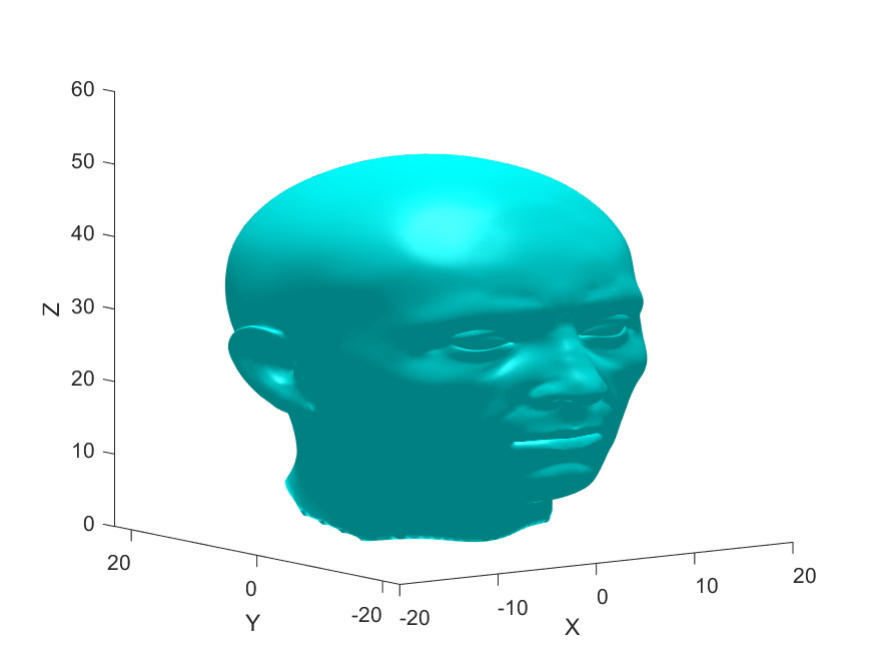}
\end{minipage}
}
\subfigure[Tangent plane]{
\begin{minipage}[b]{0.22\textwidth}
\includegraphics[width=1\textwidth]{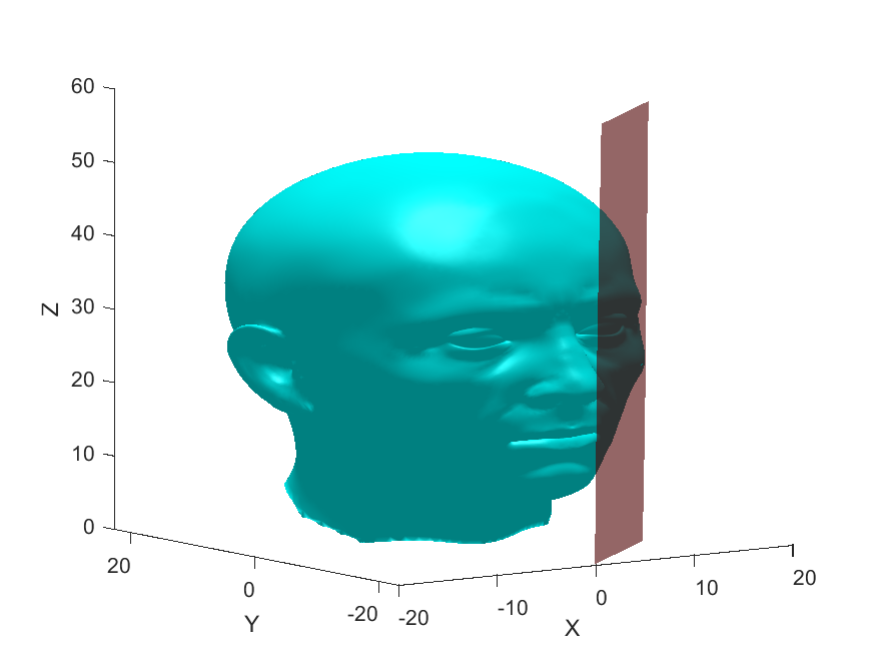}
\end{minipage}
}
\caption{Illustration of the minimum distance from intracavity lasers to the eye upon invasion}
  \label{fig:humanhead}
\end{figure}

\subsection{Eye Interaction Angle and Distance}
To model the interaction between the intracavity laser and the human head, the goal is to determine two key parameters: the distance and the incident angle between the eye and the first facial structure that interacts with the laser beam. These parameters will help us understand the depth of the head's intrusion into the laser beam path and the orientation of the laser beam relative to the eye. To achieve this, we treat the problem as a geometric analysis involving the eye and the outer tangent planes of the 3D head model. The outer tangent planes are those that touch the model's surface without \ml{intersecting its interior}, and we aim to calculate the distance between the eye and the closest tangent plane as well as the incident angle of this plane relative to the eye's normal vector.
\subsubsection{Model Presentation and Convex Hull Calculation} Let the 3D model of a portion of the head, with its height encompassing the eye, be represented as a set of vertices $P=\{\mathbf{p}_1, \mathbf{p}_2,..., \mathbf{p}_F\}$, where each point $\mathbf{p}_i = (x_i,y_i,z_i)$ is a point in $\mathbb{R}^3$ extracted from the STL model of the head. These vertices define the geometry of the model, which we use to calculate its convex hull. The convex hull of the head model ${\rm conv}(P)$ is the smallest convex polyhedron that encloses all the points in $P$, defined as
\begin{equation}
    \operatorname{conv}(P)=\left\{\sum_{i=1}^F \omega_i \mathbf{p}_i \mid \sum_{i=1}^F \omega_i=1, \omega_i \geq 0\right\},
\end{equation}
where $\omega_i$ indicates the weight. The convex hull forms the outer boundary of the head, and its faces correspond to the external tangent planes of the model. It can be obtained using a computational geometry algorithm (such as Quickhull) or a built-in MATLAB function like \textit{convhull}$(\cdot)$.
\subsubsection{Tangent Plane to the Eye's Position} Once the convex hull is computed, its faces correspond to the external tangent planes of the 3D model. Suppose the three vertices defining the plane are $\mathbf{v}_l,\mathbf{v}_m,\mathbf{v}_n$, the normal vector $\mathbf{n}_{k}$ to this plane can be computed using the cross product of two edge vectors:
\begin{equation}
    \mathbf{n}_{k} = (\mathbf{v}_m-\mathbf{v}_l)\times (\mathbf{v}_n-\mathbf{v}_l).
\end{equation}
Each face of the convex hull can be expressed by a plane equation:
\begin{equation}
    \mathbf{n}_{k} \cdot\left(\mathbf{r}-\mathbf{v}_l\right)=0,
\end{equation}
where $\mathbf{r}$ represents any point on the plane.
\subsubsection{Distance and Angle Calculation}
For each tangent plane, we compute its distance to the eye. Let the set of eye surface positions be defined as $P_e=\{\mathbf{e}_1, \mathbf{e}_2,..., \mathbf{e}_E\}$ where each point $\mathbf{e}_j = (x_j,y_j,z_j)$ represents a position on the eye surface. The distance $d_{kj}$ between the $k$-th tangent plane and the eye position $\mathbf{e}_j$ is given by:
\begin{equation}
    d_{kj} = \frac{\left|\mathbf{n}_k \cdot\left(\mathbf{e}_j-\mathbf{v}_l\right)\right|}{\left\|\mathbf{n}_k\right\|},
\end{equation}
where $|\cdot|$ denotes the absolute value operator and $||\cdot||$ represents the norm operator. The angle between the eye’s normal vector $\mathbf{n}_{e}$ (determined based on the human head model) and the $k$-th tangent plane is calculated as
\begin{equation}
    \theta=\cos ^{-1}\left(\frac{\mathbf{n}_e \cdot \mathbf{n}_k}{\left\|\mathbf{n}_e\right\|\|\mathbf{n}_k\|}\right).
\end{equation}

The process of determining the outer tangent plane for the human head is illustrated in Fig.~\ref{fig:humanhead}. We use the STL model of a typical human head, as shown in Fig.~\ref{fig:humanhead} (a). In Fig.~\ref{fig:humanhead} (b), a tangent plane with the minimum distance to the eye is highlighted. This distance, along with the angle between the plane and the eye’s normal vector, is then used to assess whether the beam will be focused by the eye.

\begin{figure}  
\centering
\subfigure[Beam focusing by eye]{
\begin{minipage}[b]{0.22\textwidth}
\includegraphics[width=1\textwidth]{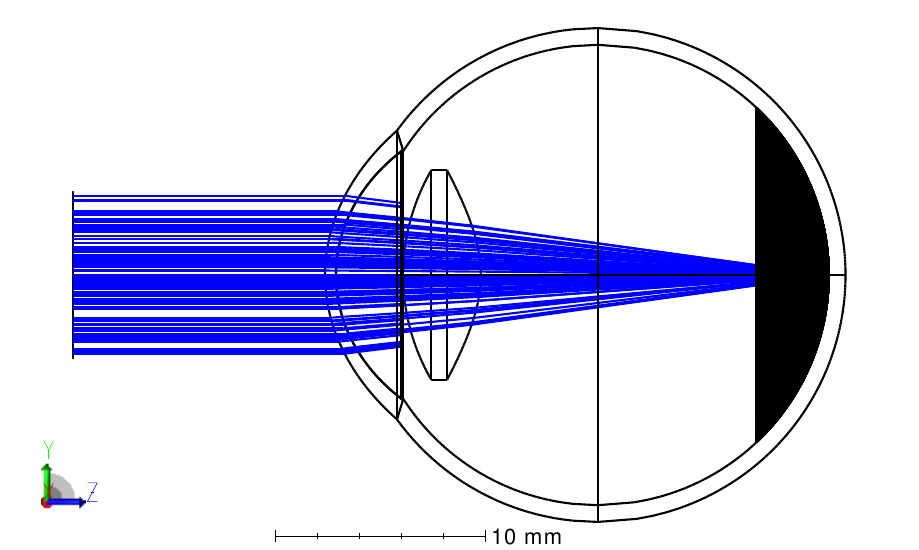}
\end{minipage}
}
\subfigure[Beam not focusing by eye]{
\begin{minipage}[b]{0.22\textwidth}
\includegraphics[width=1\textwidth]{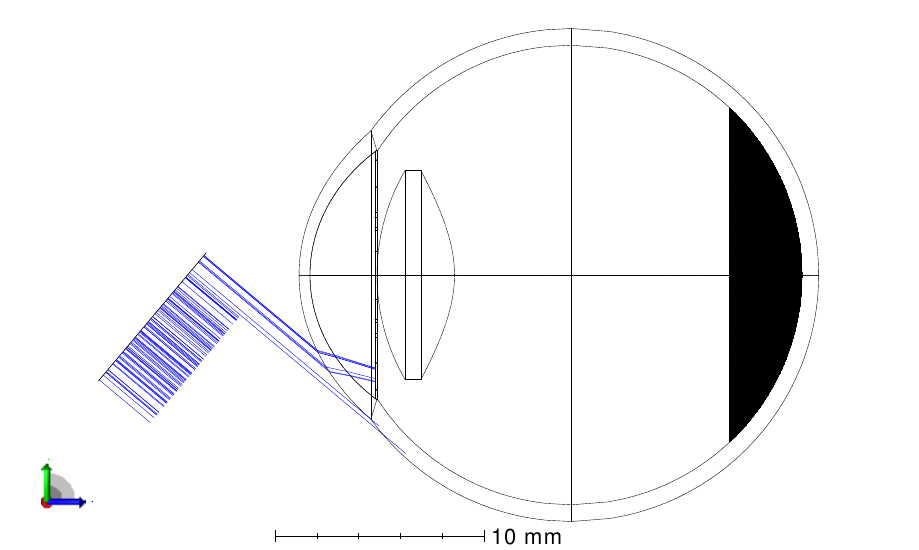}
\end{minipage}
}
 \caption{Simulation on beam focusing by eye structure in Zemax.}
  \label{fig:eyemodel}
\end{figure}

\begin{table}[!htbp]
\centering
\caption{~Parameters of DCCL-WPT Simulation}
\vspace{.7em}
\begin{tabular}{ccc}
\hline
\textbf{Parameters}&\textbf{Symbols}&\textbf{Value}\\
\hline
\text{CRR radius}&$r_{\rm c}$&$6$ mm\\
\text{Focal length of L1$\sim$L4}&$f_1\sim f_4$&$15$ mm\\
\text{Reflectivity of CRR1}&$R_1$&$1$\\
\text{Reflectivity of CRR2/3}&$R_2/R_3$&$0.5$\\
\text{Reflectivity of CRR4}&$R_4$&$0.1$\\
\text{intracavity laser wavelength}&$\lambda$ & $1064$ nm\\
\text{Gain medium radius}&$r_{\rm g}$&$6$ mm\\
\text{Gain medium length}&$l_{\rm g}$&$1$ mm\\
\text{Medium saturated intensity}&$I_{\rm s}$&$1.20\times10^{7}$\\
\text{Excitation efficiency}&$\eta_{\rm c}$&$0.72$\\
\text{Radius of invasion plane}&$r_{\rm inv}$ & $12$ mm\\
\text{Free-space cavity length}&$L_{BF}+L_{FB}$ & $5$ m\\
\text{Sampling number}&$M$&$4096$\\
\text{Computation window expand factor}&$G$ & $3$\\
\text{PV's responsivity}&$\rho$ & $0.6$ A/W\\
\text{Reverse saturation current}&$I_0$ & $0.32~\mu$A\\
\text{Number of cells in PV}&$n_s$ & $1$\\
\text{Diode ideality factor}&$n$ & $1.48$\\
\text{Temperature}&$T$ & $298$K\\
\text{Shunt resistance}&$R_{\rm sh}$ & $53.82~\Omega$\\
\text{Series resistance}&$R_{\rm s}$ & $372~m\Omega$\\
\hline
\label{t:DCCLPara}
\end{tabular}
\end{table}

\subsection{Ray Tracing for Beam Focus Analysis}
To evaluate whether the intracavity laser beam is focused onto the retina, we establish a simulation framework in \ml{Ansys Zemax OpticStudio} using its built-in 3D eye model. The eye model consists of multiple anatomical components, including the cornea, lens, vitreous humor, and retina, each with their respective refractive indices and optical functions. The cornea and lens together form the primary focusing mechanism of the eye, bending incident light towards the retina.

In our simulation, we introduce a parallel light source to mimic the intracavity laser beam. This light source allows precise control over both the incident angle and entry position, enabling us to analyze different laser entry conditions. The key steps in the ray-tracing process include: i) initial refraction by the cornea and the extent of refraction depends on the beam’s incidence angle and position relative to the corneal curvature; ii) focusing by the lens which exhibits a gradient refractive index, dynamically adjusting its focal power; iii) propagation through the vitreous humor and if the beam remains collimated or diverging at this stage, it will not be focused onto the retina.

By systematically adjusting the beam incidence conditions based on the distance and angle determined in Sec III-A, we verify whether the laser will be focused onto the retina. If the traced rays fail to converge at the retinal surface, it confirms that under the given conditions, the intracavity laser does not pose a retinal hazard. An illustrative simulation is shown in Fig.~\ref{fig:eyemodel}, demonstrating that the eye does not focus the beam with certain incident angles.

\begin{figure*}  
\centering
\subfigure[DCCL vs. DSCL]{
\begin{minipage}[b]{0.31\textwidth}
\includegraphics[width=1\textwidth]{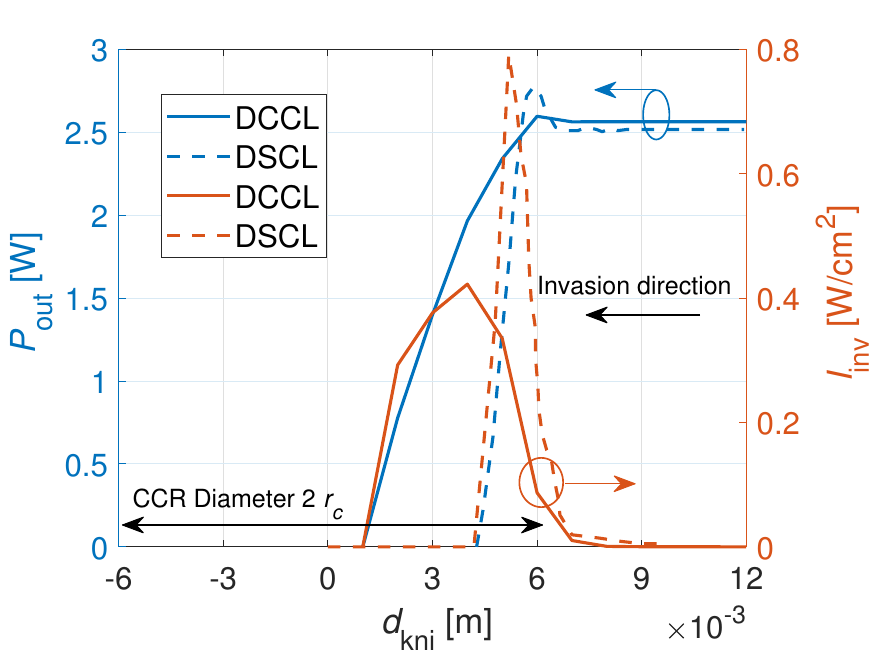}
\end{minipage}
}
\subfigure[Varying fill factors]{
\begin{minipage}[b]{0.31\textwidth}
\includegraphics[width=1\textwidth]{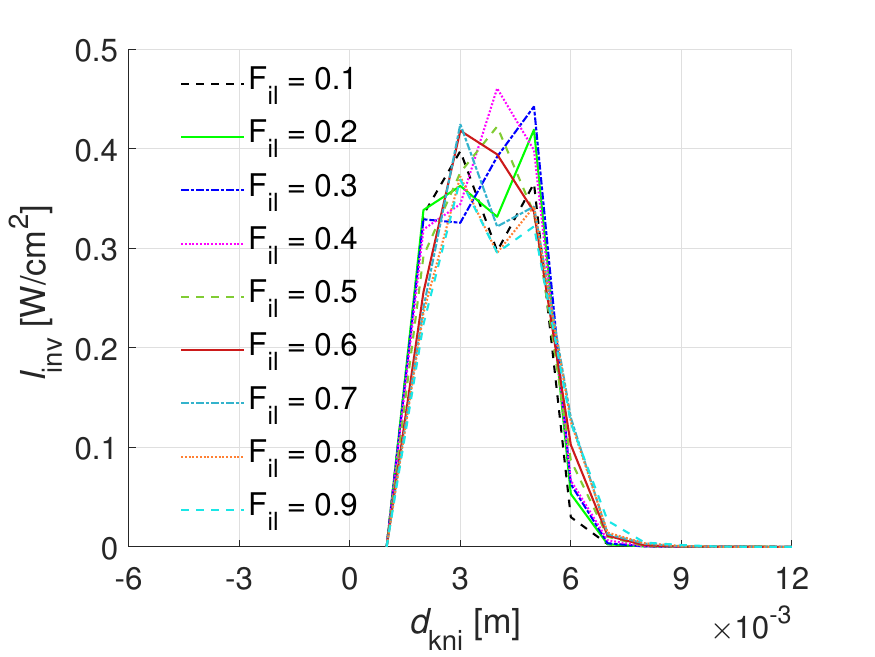}
\end{minipage}
}
\subfigure[Varying moving angles]{
\begin{minipage}[b]{0.31\textwidth}
\includegraphics[width=1\textwidth]{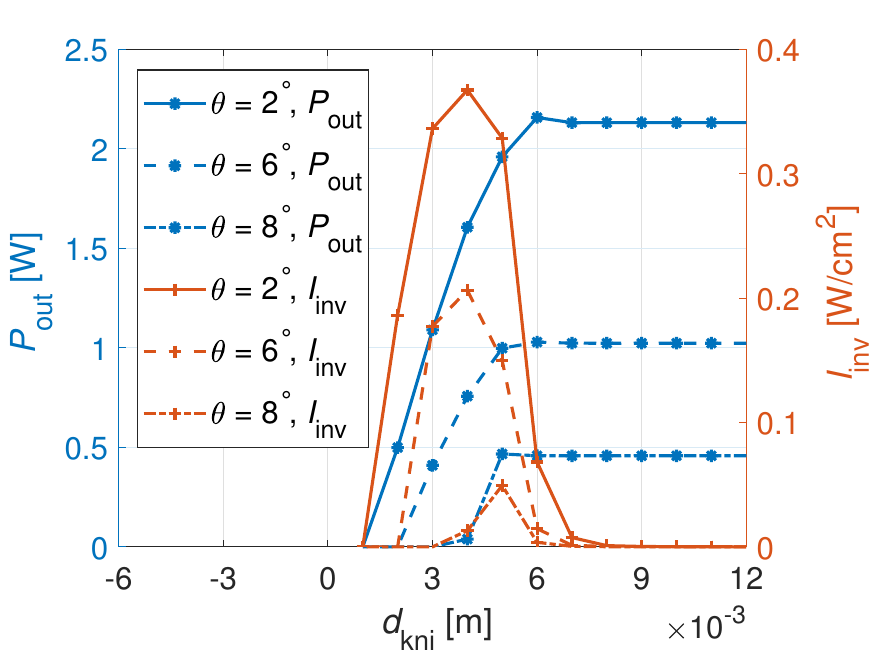}
\end{minipage}
}
\caption{Skin safety analysis for DCCL system under comparison with DSCL, varying fill factors, and moving angles.}
\label{f:skinsafety}
\end{figure*}

\section{Simulation Results and Discussions}
This section provides a comprehensive numerical analysis for the safety assessment of the DCCL-WPT system. First, we compare its safety limits with those of the DSCL-WPT system. Then, we conduct a numerical evaluation of eye safety. Next, we analyze the system's sensitivity to small intruding objects. The achievable wireless charging power under safety constraints is estimated subsequently. Finally, we discuss key findings and outline directions for future research.
\subsection{Parameters}
The parameters used for the safety assessment of the DCCL-WPT system correspond to those required for calculating the irradiance on the intruding object within the free-space cavity. Therefore, we define key parameters for the DCCL structure, including CCR radius, focal length, gain parameters, and cavity length, as well as parameters for foreign objects such as size and position. To ensure a direct comparison between the DCCL and DSCL based WPT systems, the resonant cavity parameters, such as reflector radius and cavity length, are set to match those of the DSCL based WPT system, following the safety analysis in~\cite{basicsafety}. Furthermore, we investigate safety limits under varying mobility conditions, selecting FOV parameters based on~\cite{basicMobility}. The gain medium is a thin-disk Nd:YVO$_{4}$, with parameters derived from measurements in~\cite{hodgson2005laser}. For intruding objects, the knife-edge parameters are taken from~\cite{basicsafety}. Besides, as shown in \cite{mobilityEnhanced,mobilitySMIPT}, zero-padding and appropriate sampling are necessary for FFT implementation. For power conversion from laser beam to electrical energy, we choose an InGaAsP-based PV, of which the key parameters are obtained from \cite{PVparaLONG}. A detailed list of parameters is provided in Table~\ref{t:DCCLPara}.

\begin{figure}
    \centering
    \includegraphics[width=0.5\linewidth]{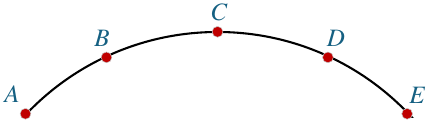}
    \caption{Illustration of selecting typical points on the eye surface.}
    \label{fig:eyepoint}
\end{figure}

\subsection{Skin Safety Enhancement of DCCL-WPT}
In the DSCL-WPT system's skin-safety analysis, the output beam power from the receiver is set to $2.5$W~\cite{basicsafety}. For a fair comparison, we determine the parameters required to achieve an equivalent output power in the DCCL-WPT system using our proposed numerical analysis method. The resulting output power is $P_{\rm out} = 2.5637$W, with a corresponding input power of $P_{\rm in}=86.6$W. As in~\cite{basicsafety}, the object is assumed to move perpendicularly to the optical axis, intruding from a distance of $2r_c=0.012$m to $-r_c=-0.006$m, until the reflectors forming the resonant cavity are fully blocked. The intrusion position $d_{\rm kni}$ represents the vertical coordinate of the intruding object relative to the optical axis.

Fig.~\ref{f:skinsafety} presents the skin-safety analysis for the DCCL-WPT system, comparing it with DSCL-WPT and evaluating the effects of different intruding positions within the free-space cavity. In Fig.~\ref{f:skinsafety} (a), we depict the output power $P_{\text{out}}$ and irradiance $I_{\text{inv}}$ on intrusion object surface as a function of the vertical intrusion position $d_{\text{kni}}$. The results show that the DCCL-WPT system maintains a higher output power while limiting the irradiance on intruding objects, thereby enhancing safety compared to the DSCL-WPT. The safety improvement in terms of maximum $I_{\text{inv}}$ reduction is nearly twofold.
To verify that intruding object at any horizontal position within the free-space cavity meets the skin safety requirements, the subplot Fig.~\ref{f:skinsafety} (b) examines the impact of different $F_{il}$ on $I_{\text{inv}}$. As $F_{il}$ changes, the position of maximum irradiance shifts slightly; however, all cases remain within the skin-safe threshold. To verify whether the system remains safe under dynamic conditions, we analyze the impact of different transceivers' angles $\theta$ on $P_{\text{out}}$ and $I_{\text{inv}}$ as in Fig.~\ref{f:skinsafety} (c). As $\theta$ increases, both the output power and intrusion irradiance decrease, aligning with the transmission channel variations under mobility. Nevertheless, the results confirm that the system remains skin-safe during movement.

\begin{table}[]
\caption{Angles between the tangent plane normal and the corneal surface where the separation is less than $3$ mm}
\label{tab:eyeangle}
\centering
\begin{tabular}{l|l|l|l|l}
\hline
Distance {[}mm{]}   & 2.38874 & 2.7443 & 2.87897 & 2.94929 \\ \hline
Angle {[}$^{\circ}${]} & 141.557 & 140.46 & 139.076 & 138.534 \\ \hline
\end{tabular}
\end{table}

\begin{figure*}  
\centering
\subfigure[Point A]{
\begin{minipage}[b]{0.18\textwidth}
\includegraphics[width=1\textwidth]{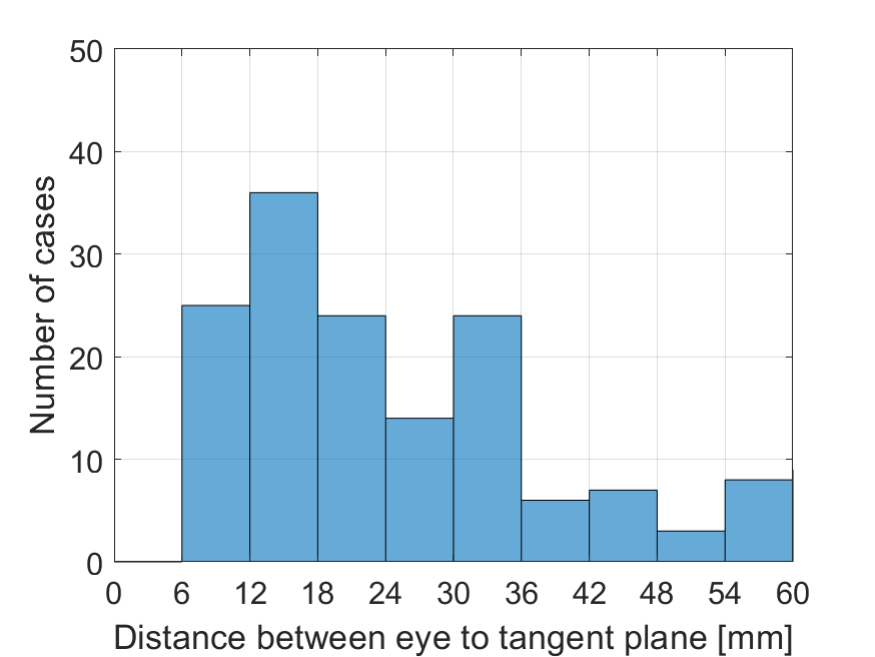}
\end{minipage}
}
\subfigure[Point B]{
\begin{minipage}[b]{0.18\textwidth}
\includegraphics[width=1\textwidth]{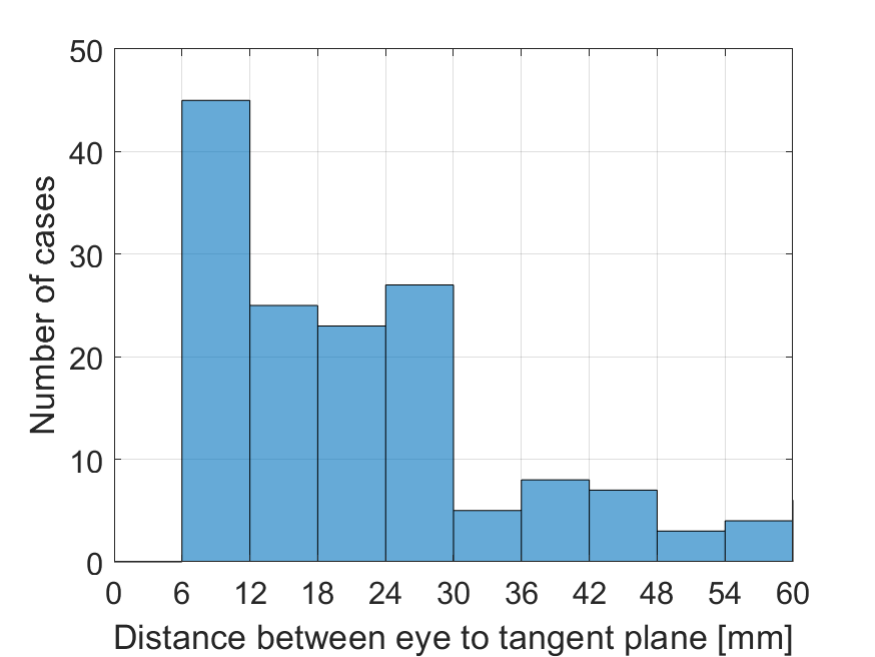}
\end{minipage}
}
\subfigure[Point C]{
\begin{minipage}[b]{0.18\textwidth}
\includegraphics[width=1\textwidth]{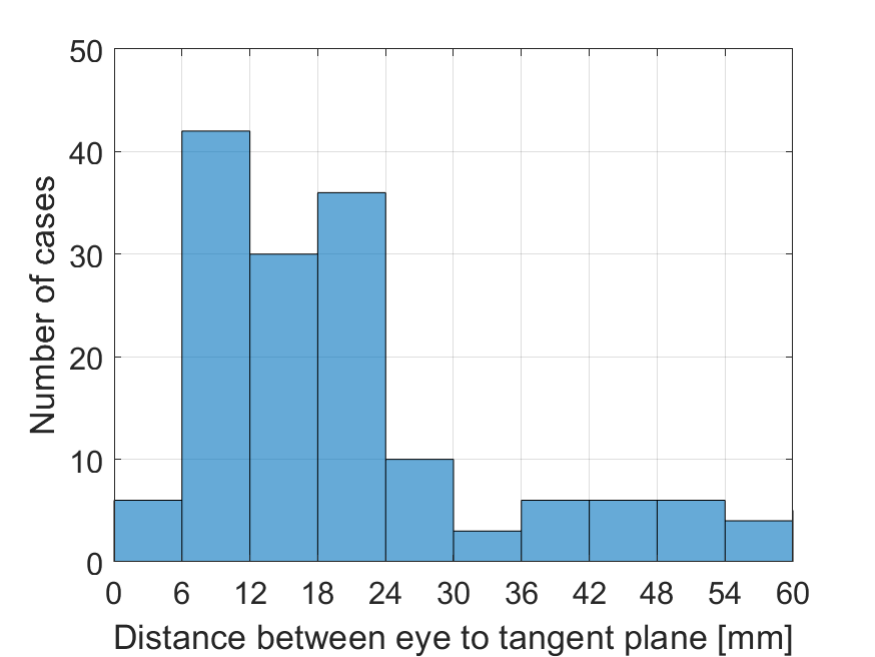}
\end{minipage}
}
\subfigure[Point D]{
\begin{minipage}[b]{0.18\textwidth}
\includegraphics[width=1\textwidth]{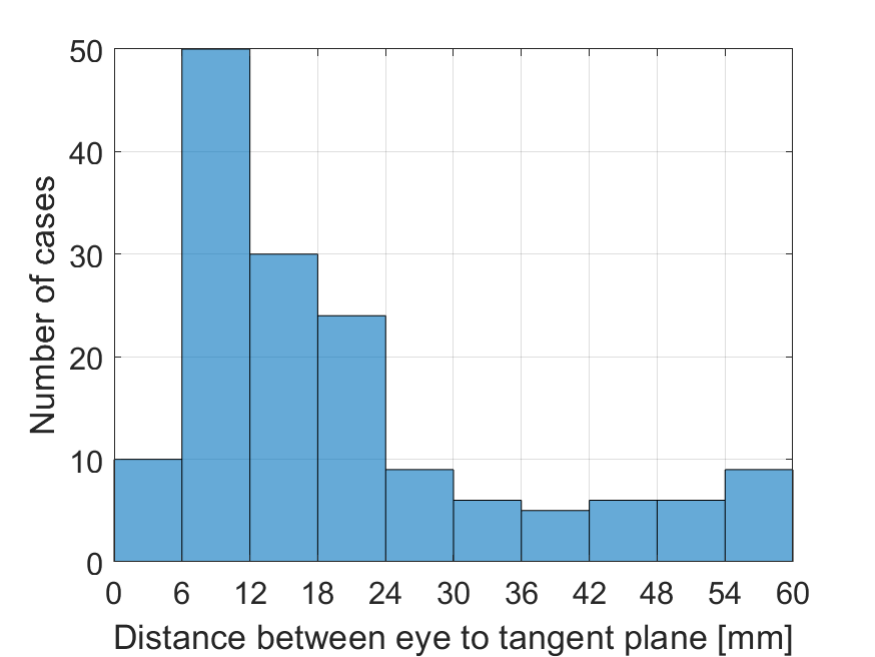}
\end{minipage}
}
\subfigure[Point E]{
\begin{minipage}[b]{0.18\textwidth}
\includegraphics[width=1\textwidth]{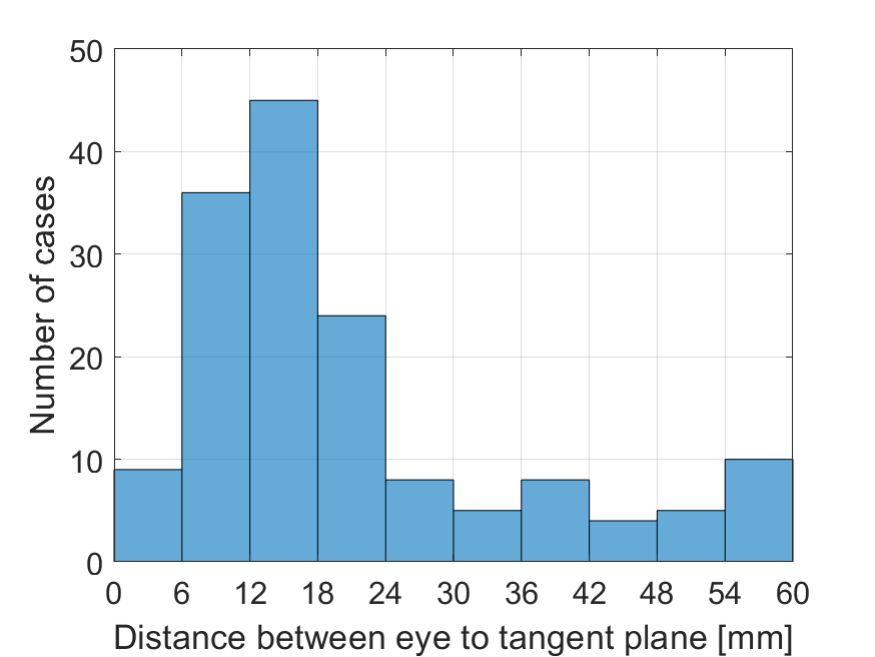}
\end{minipage}
}
\caption{Distances between all tangent planes to (a) point A, (b) point B, (c) point C, (d) point D, (e) point E on the eye surface.}
\label{f:stochastic}
\end{figure*}

\begin{figure*}  
\centering
\subfigure[$P_{\rm out}$ vs. $d_{\rm kni}$]{
\begin{minipage}[b]{0.31\textwidth}
\includegraphics[width=1\textwidth]{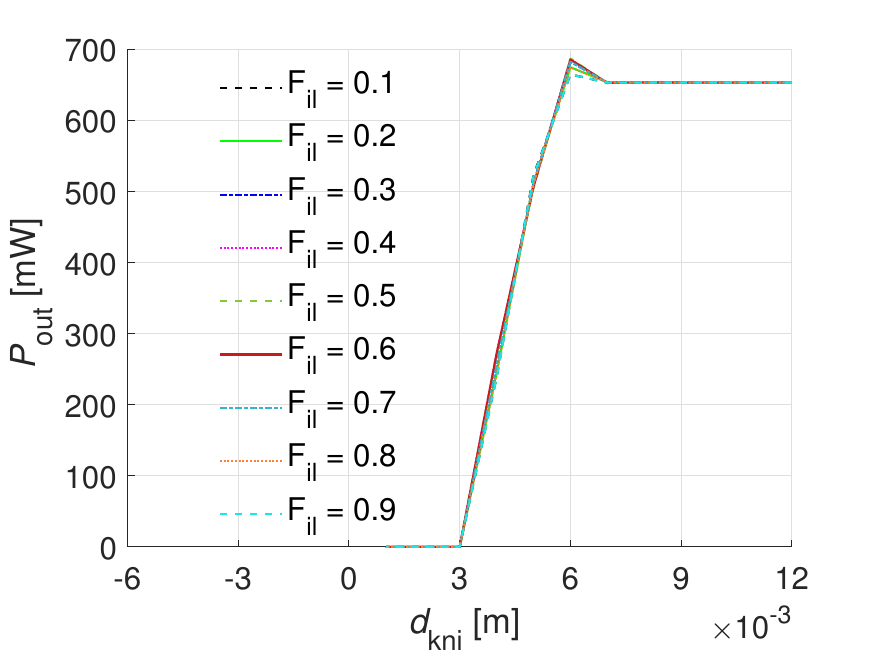}
\end{minipage}
}
\subfigure[$I_{\rm inv}$ vs. $d_{\rm kni}$]{
\begin{minipage}[b]{0.31\textwidth}
\includegraphics[width=1\textwidth]{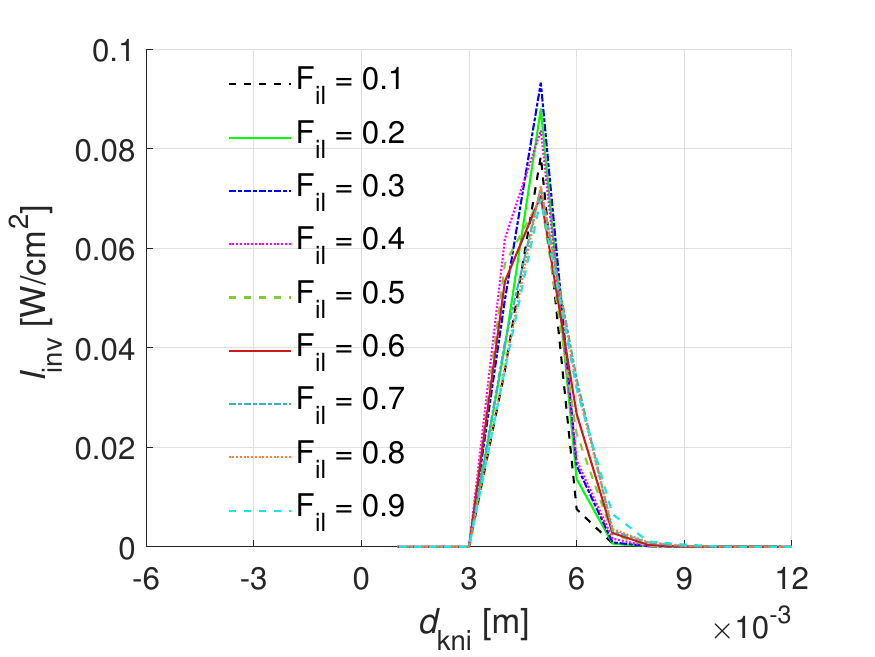}
\end{minipage}
}
\subfigure[Eye focusing simulation]{
\begin{minipage}[b]{0.31\textwidth}
\includegraphics[width=1\textwidth]{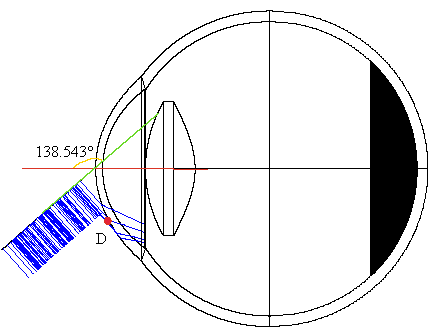}
\end{minipage}
}
\caption{Eye safety analysis for DCCL system.}
\label{f:eyesafety}
\end{figure*}
\subsection{Eye Safety in DCCL-WPT}
We select different positions on the front surface of the eye to calculate the distance between the eye and the outer tangent planes of the human head. These positions represent potential beam paths and the points where the beam may make contact with the eye. For simplicity, we select the arc passing through the corneal center, which can be represented by a circular function. We choose five points along this arc to represent potential beam contact points with the eyeball. This allows us to analyze whether the beam will be interrupted before reaching the eyeball and whether it will be focused upon contact. From the human head model as shown in Fig.~\ref{fig:humanhead} (the scale ratio between the dimensions shown on the coordinate system and the actual size of the human head is 200:1), we first extract the coordinate of the start, center, and end point of this arc $A,C,E$ as $(3.9037, -14.6738, 29.6959)$, $(6.46821, -15.0006, 30.2467)$, $(8.60596, -13.6036, 29.5649)$. The arc function is then calculated as $(x-1.3217)^2+(y-0.762)^2=1.499^2$. Finally, we select two additional points $B(5.124,-15.068,30.121)$ and $D(8.232,-14.057,29.827)$ to ensure that the distance between each pair of points is equal, as depicted in Fig.~\ref{fig:eyepoint}.

For the five selected points (A–E) along the arc passing through the corneal center, Fig.~\ref{f:stochastic} presents histograms of the distances between the eye and the tangent planes of surrounding head structures. In this context, the distance represents the separation between the first contact point of the intracavity laser with the surrounding head structures and its subsequent contact with the corneal surface. A larger distance indicates that the laser beam must traverse a longer path through intervening structures before reaching the cornea, making it more vulnerable to interruption by these structures during human movement or partial intrusion.
It is observed that for points A and B, no instances exhibit distances smaller than $6$ mm (CCR surface radius), suggesting a lower probability of immediate beam contact with the cornea. For points C, D, and E, although some cases show distances below $6$ mm, these instances remain rare. Based on this analysis, two key points need to be verified: i) the intrusion depth at which the intracavity beam is cut off while ensuring corneal safety, and ii) in cases where the beam transmission does not cease before reaching the cornea (as shown in Fig.~\ref{f:stochastic}), whether the beam will be focused by the eye despite the shorter distance.

Following a similar methodology as in Sec. IV. B, we further investigate the irradiance and output power of the DCCL-WPT system under the scenario where a human eye intrudes into the intracavity laser. The input power $P_{\rm in}$ is set to $85.35$ W. Fig.~\ref{f:eyesafety} (a) and (b) shows the the output power $P_{\text{out}}$ and invasion irradiance $I_{\text{inv}}$ as functions of the vertical invasion position $d_{\text{kni}}$, under various fill factors $F_{il}$. As in Fig.~\ref{f:eyesafety} (a), as $d_{\text{kni}}=12$ mm (i.e., $6$ mm away from the cavity edge), system output power reaches over $650$ mW. As the intruding object approaches the cavity edge, $P_{\rm out}$ first increases, then decreases as the object continues to intrude, eventually dropping to zero at $d_{\text{kni}}=3$ mm, indicating that the intracavity laser has been fully interrupted. Correspondingly, Fig.~\ref{f:eyesafety} (b) illustrates the irradiance on the intruding object, which rises and then falls to zero as the intrusion depth reaches $3$ mm. Throughout the entire process, the irradiance $I_{\text{inv}}$ remains below $0.1$ W/cm$^2$. These results confirm that once the intrusion depth reaches $3$ mm, the intracavity laser is shut down, and during the invasion process the irradiance remains below the MPE threshold for cornea safety with $650$ mW output power. 

To ensure safety, we then identify all situations where the distance between the tangent plane of the human face and the corneal surface is less than $3$ mm, implying a potential risk of corneal contact before the laser is cut off. Among all configurations analyzed in Fig.~\ref{f:stochastic}, only the Point D configuration contains cases with such close proximity. Table~\ref{tab:eyeangle} lists the corresponding angles between the tangent plane normal and the corneal normal vector for these cases. The minimum observed angle is $138.534^{\circ}$, which we use in the eye model in Ansys Zemax OpticStudio to verify whether such an incident angle can result in laser beam focusing onto the retina. Fig.~\ref{f:eyesafety} (c) demonstrates that the beam is not focused by the eyeball under this condition. The contact point between the beam and the cornea corresponds to the predefined point D, and the beam's incident angle, tilted along the y-axis, is set to $48.534^{\circ}$. With a system output power of $650$ mW, even if the eye approaches and contacts the intracavity laser beam, the beam will not be focused onto the retina. Irradiance analysis further confirms that corneal safety can be guaranteed.

\begin{figure}
    \centering
    \includegraphics[width=0.7\linewidth]{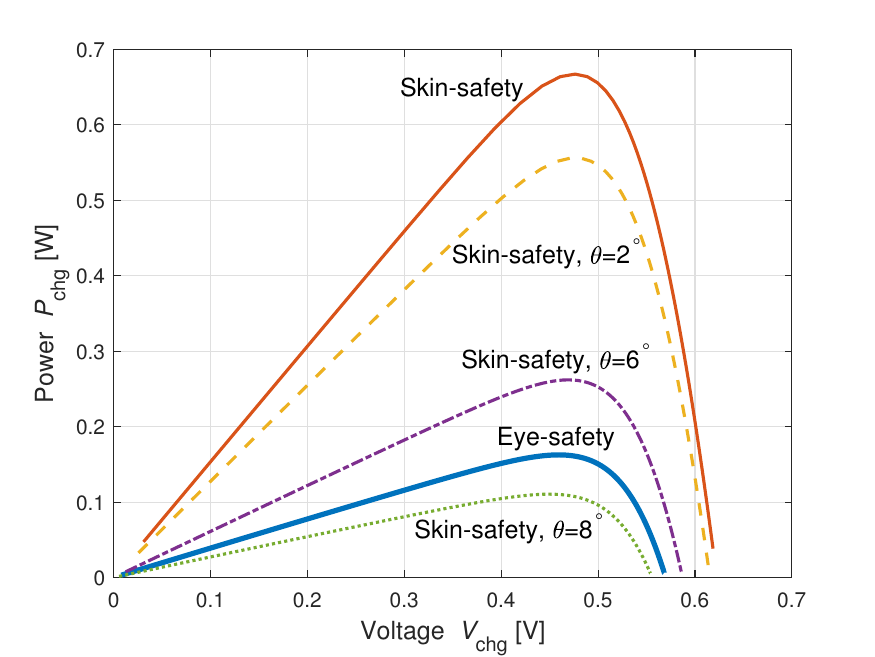}
    \caption{P-V curves for DCCL-WPT system under various safety constraints.}
    \label{fig:pvcurve}
\end{figure}

\begin{figure}
    \centering
    \includegraphics[width=0.7\linewidth]{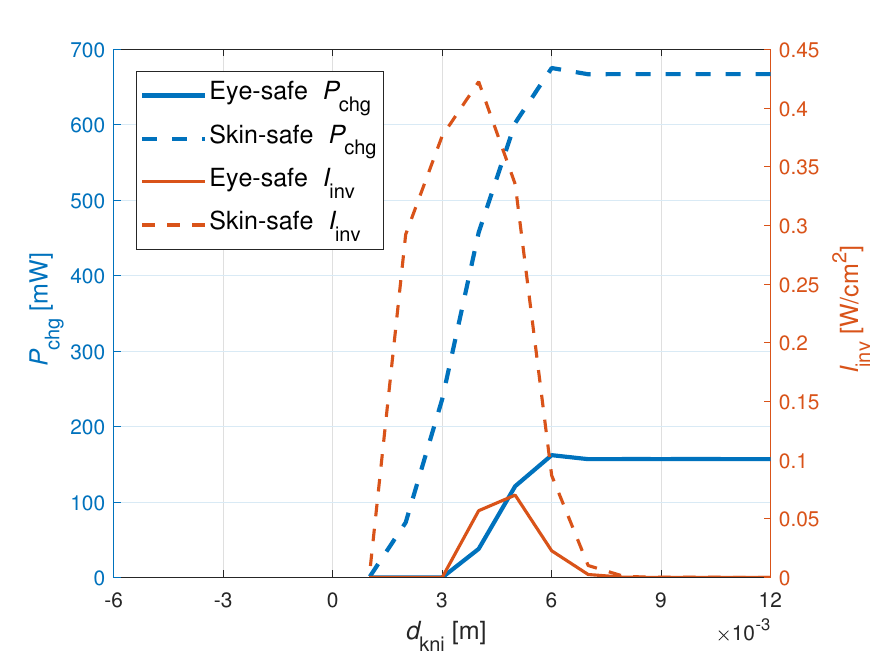}
    \caption{Skin-safe and eye-safe wireless charging power and corresponding irradiance inside the cavity.}
    \label{fig:WPTsafety}
\end{figure}

\subsection{Wireless Charging Power under Safety Constraints}
Fig.~\ref{fig:pvcurve} illustrates the power–voltage (P–V) characteristics of the proposed DCCL-WPT system under various safety constraints, including eye-safety and skin-safety conditions with different transceiver angles $\theta$. Due to the distinct allowable output beam powers under these conditions, as shown in Figs.~\ref{f:skinsafety} and~\ref{f:eyesafety}, the corresponding PV-converted wireless charging power varies accordingly. By employing MPPT, the peak wireless charging power and the corresponding optimal load resistance $R_{\rm L}$  can be obtained for each configuration.
With the optimal $R_{\rm L}$, Fig.~\ref{fig:WPTsafety} presents the skin-safe and eye-safe wireless charging power along with the corresponding intracavity irradiance $I_{\mathrm{inv}}$ as functions of the intruding distance $d_{\mathrm{kni}}$. It is observed that the skin-safe configuration enables significantly higher peak charging power, exceeding $600$~mW, whereas the eye-safe configuration yields a peak power of approximately $150$~mW at a transceiver distance of $5$~m. Furthermore, more than $100$~mW of wireless charging power can be maintained within a $16^\circ$ self-alignment moving FoV. Meanwhile, the intracavity irradiance profiles highlight the critical safety threshold regions, delineating operational boundaries to ensure compliance with regulatory limits. Lastly, with $5$ m transceiver distance,

\subsection{Sensitivity to Small Object Intrusion}
\begin{figure}
    \centering
    \includegraphics[width=0.7\linewidth]{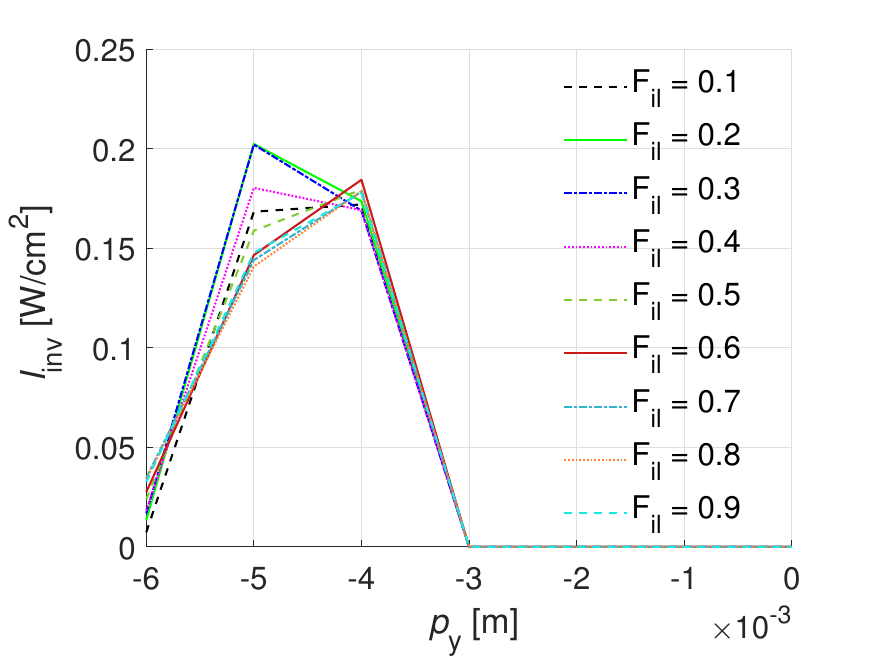}
    \caption{Irradiance on hair at different positions inside the cavity.}
    \label{fig:smallMPE}
\end{figure}

With the boundary condition given in Eq.~\eqref{eq:hairbound}, we analyze the irradiance changes when a small object (hair) intrudes into the cavity. The hair, with dimensions $l_{\rm hr} = 24$ mm along the $x$-axis and $w_{\rm hr} = 70~\mu$m along the $y$-axis, is positioned at $p_x = -12$ mm while varying $p_y$ to observe its effect. System input power $P_{\rm in}$ is set to $85.35$ W, consistent with the value employed in the aforementioned eye safety verification tests. In Fig.~\ref{fig:smallMPE}, as the hair moves from the cavity edge ($p_x = -6$ mm) toward the center, the irradiance initially increases due to the higher power density near the cavity's central region. However, when the hair reaches 
$p_x = -2$ mm, i.e., halfway to the center, the irradiance drops abruptly, indicating intracavity laser beam interruption. The sharp decline demonstrates that even a minor intrusion significantly disrupts the cavity's resonant field, avoiding the expectation of further irradiance increase closer to the center, which highlights the sensitivity of the DCCL-WPT system to small-object intrusions.

\subsection{Discussions}
Our comprehensive safety analysis of the DCCL system yields three key findings. 
First, when operating at an output power of $2.5$ W, the DCCL-WPT system satisfies the skin safety requirement ($< 1$ W/cm$^2$). Moreover, the irradiance is reduced by half compared to the DSCL structure under equivalent power conditions. We also verify that this safety advantage remains stable even under dynamic conditions, i.e., both output power and local irradiance on intruding objects decrease with receiver movement.
Second, for eye safety, the analysis of the DCCL-WPT system can be simplified to corneal exposure. At an output power of $650$ mW, the measured irradiance consistently stays below the $0.1$ W/cm$^2$ corneal safety threshold. Even in worst-case scenarios where the laser beam reaches the eye prior to interruption, the incident angle prevents retinal focusing, and the irradiance is below safe exposure limits. Third, the DCCL-WPT system demonstrates high sensitivity to small-object intrusions, such as hair. When intruding objects approach the optical axis, where power density is highest, the resonant field is rapidly disrupted, prompting immediate laser shutdown. This intrinsic response mechanism effectively prevents localized overexposure at critical positions, reducing the risk of damage to both biological tissues and small intruding objects. Notably, the proposed DCCL-WPT system can achieve over $600$~mW of wireless charging power under skin-safe conditions and approximately $150$~mW under eye-safe constraints at a $5$~m transceiver distance, while still maintaining over $100$~mW within a $16^\circ$ self-alignment moving field of view, underscoring its suitability for practical, mobile, and safety-compliant energy transfer applications.

\section{Conclusions}
\label{sec:Conclusion}
This study presents a comprehensive safety assessment of distributed coupled-cavity laser-based wireless power transfer (DCCL-WPT) systems under three critical scenarios: skin exposure, eye interaction, and small-object intrusion. By integrating beam diffraction modeling and gain-loss dynamics, we demonstrate that DCCL-WPT achieves over $600$ mW of wireless charging at $5$ m under skin-safe conditions, while reducing irradiance on large-scale intrusions by nearly $50$\% compared to single-cavity designs. Eye safety is ensured via anatomical and optical analysis, showing that even at $650$ mW output ($1064$ nm), the beam does not reach the retina and corneal exposure remains below $0.1$ W/cm\textsuperscript{2}, enabling $150$ mW power delivery under eye-safe conditions. Furthermore, the system exhibits high sensitivity to small-object intrusion, providing both passive protection and early hazard detection. These findings support the feasibility of advancing laser-based systems from high-speed communication toward safe and practical WPT.

Despite the promising safety characteristics demonstrated through modeling and simulation, experimental validation of intracavity laser safety remains challenging due to the closed nature of the resonant structure and the difficulty of directly measuring irradiance distributions within the cavity. Future work should explore novel indirect measurement techniques or surrogate testing platforms to verify the predicted safety margins under realistic operating conditions. In addition, further efforts can be directed toward enhancing safety through artificial intelligence (AI)-based strategies, to further improve DCCL's resilience in dynamic, multi-user scenarios.
\bibliographystyle{IEEEtran}
\bibliography{Reference}
\end{document}